\documentclass[twocolumn,twoside,preprintnumbers,amsmath,amssymb,pacs,floatfix]{revtex4}
\usepackage{epsfig}
\usepackage{graphicx}% Include figure files
\usepackage{dcolumn}% Align table columns on decimal point
\usepackage{bm}% bold math
\usepackage{fancyhdr}
\usepackage{pslatex}
\newcommand{\be}{\begin{eqnarray}}
\newcommand{\ee}{\end{eqnarray}}
\def\beq{\begin{equation}}
\def\slashiii#1{\setbox0=\hbox{$#1$}#1\hskip-\wd0\hbox to\wd0{\hss\sl/\/\hss}}
\def\eeq{\end{equation}}
\def\SB{S$\chi$SB}
\def\fm{\,\mathrm{fm}}
\begin{document}

\title{{\Large Is the chiral phase transition induced by a metal-insulator transition?  }}

\author{Antonio M. Garc\'{\i}a-Garc\'{\i}a}

\affiliation{Physics Department, Princeton University, Princeton,
New Jersey 08544, USA}

\affiliation{The Abdus Salam International Centre for Theoretical
Physics, P.O.B. 586, 34100 Trieste, Italy}
\author{James C. Osborn}
\affiliation{Physics Department \& Center for Computational Science,
 Boston University, Boston, MA 02215, USA}

\received{on 24 March, 2006}

\begin{abstract}
We investigate the QCD Dirac operator with gauge configurations
given by a liquid of instantons 
 in the region of temperatures about the chiral phase transition. Both the 
quenched and unquenched cases are examined in detail.  
We present evidence of a localization transition in the low lying modes of the
 Dirac operator around the same temperature as the chiral phase transition.
Thus both level statistics and eigenvectors of the QCD Dirac operator 
at the chiral phase transition have 
similar properties than those of a disordered conductor at 
the metal-insulator transition. 
This strongly suggests the phenomenon of Anderson localization 
(localization by destructive quantum interference)
 is the leading physical mechanism in the restoration of the   
chiral symmetry.
Finally we argue that our findings are not in principle restricted to the ILM
 approximation and may also be found in lattice simulations.


\end{abstract}

\maketitle

\thispagestyle{fancy}
\setcounter{page}{0}

The
spontaneous breaking of the approximate chiral symmetry (\SB{}) and its eventual restoration at finite temperature is one of the most important features of the strong interactions.
The order parameter associated to this symmetry breaking is the chiral 
 condensate, $\langle {\bar \psi} \psi \rangle$.
By definition, the condensate is nothing but a quark loop in momentum
 space. 
Thus it should naively vanish as the quark mass goes to zero.
In nature the lightest quarks are not massless so a nonzero 
condensate is expected even in a free theory.
%However as a result of the strong non-perturbative color interactions
% the condensate acquires a nonzero value.
%Though this is not the major contribution to the condensate.
However the small quark bare mass can only account for a very small percentage
of the chiral condensate, the rest has its origin in the strong non perturbative color interactions of QCD.
QCD models whose gauge configurations
are given by an interacting liquid of instantons \cite{shuryak}
provide an adequate theoretical framework to understand \SB.
Thus, based on the semiclassical picture of a QCD vacuum 
 dominated by instantons, it has been suggested  
\cite{chisbinst}
  that the
\SB{} in QCD and  the phenomenon of conductivity  
 may have similar physical origins.
Conductivity in a disordered sample is produced by electrons that although
initially bound to an impurity may become delocalized by orbital overlapping
 with nearby impurities. 
Similarly, in the QCD vacuum, the zero modes of the Dirac operator though 
initially bound to an instanton may get delocalized due to  
the strong overlap with other instantons. 
As a consequence, the chiral condensate becomes nonzero and
chiral symmetry is spontaneously broken. 
In this paper we provide evidence that
 these analogies can be extended to describe the 
 chiral phase transition in QCD at finite temperature. 
Specifically, in the context of an Instanton Liquid Model (ILM),
we show that around the temperature that the chiral 
phase transition occurs the low lying eigenmodes
 of the QCD Dirac operator undergo an Anderson transition (AT), 
%A
namely,
 a localization-delocalization transition, characterized by 
 multifractal eigenstates and critical statistics \cite{anderson}.
% For a more detailed account 
%we refer to Ref. \cite{aj2}.  
    
For the sake of clarity we
 provide a brief summary of the   
properties of a disordered system at an AT.
We recall that in three and higher dimensions there exists
 a mobility edge separating localized from delocalized states.  
In the delocalized region, eigenfunctions are extended through the sample and
  level statistics are described by random matrix theory (RMT).  
In the opposite limit, eigenfunctions
 are exponentially localized and the spectral correlations are 
described 
by Poisson statistics.       
Around the mobility edge
 eigenstates are multifractal \cite{aoki} 
and 
 level statistics are described by critical statistics \cite{kravtsov97}. 
Typical features of critical statistics 
include scale invariant spectrum \cite{sko},
 level repulsion and asymptotically linear number 
 variance.

Thus the level spacing distribution, $P(s)$ (the probability of having two eigenvalues 
at a distance $s$) goes to zero as the spacing $s$ separation does,
$P(s) \rightarrow 0$ $s \rightarrow 0 $. The number variance
 $\Sigma^2(\ell)=\langle (N_\ell -\langle N_\ell \rangle)^2 \rangle
  \sim \chi \ell$
 for $\ell \gg 1$ ($N_\ell$ is the number of eigenvalues in an interval of
length $\ell$)
 is asymptotically linear, as for an insulator ($\chi = 1$), but with a
 slope $\chi < 1$ ($\sim 0.27$ for a 3D Anderson model \cite{chiat}).
Below we show that at the chiral restoration temperature  
both eigenvalues and eigenvectors of the 
QCD Dirac operator have similar properties.
\section{Instanton Liquid Model}
In this section we highlight the main features of the QCD vacuum model 
based on instantons to be used in this paper.

Instantons \cite{polyakov} are
classical solutions of the Yang-Mills equations of motion which minimize the
action in Euclidean space. They are believed to be the leading
semiclassical contribution to the bosonic part of the
QCD path integral.  However the
construction of a consistent QCD vacuum based on instantons faces
serious technical difficulties.  Exact multi-instanton solutions are
hard to obtain since the Yang-Mills equations of motion for QCD are
nonlinear and therefore a superposition of single instanton
contributions is not itself a solution.  Additionally quantum
corrections may spoil the semiclassical picture implicitly assumed of
a QCD vacuum composed of instantons well separated and weakly
interacting.  These problems have been overcome either by invoking
variational principles \cite{diakonov1} or by phenomenologically fixing
certain parameters of the instanton ensemble.  The latter case,
usually referred to as the instanton liquid model (ILM)
\cite{shuryak} (for a modern review see \cite{SS97}),
 yields accurate estimates of vacuum condensates and
hadronic correlation functions \cite{SSV} for most light hadrons just
by setting the mean distance between instantons to be ${\bar R}\approx 1
\fm$ (which corresponds to a density $N/V \sim 1\fm^{-4}$) 
 and the mean instanton size ${\bar \rho} \approx 1/3 \fm$.
Lattice simulations have also supported the picture
 of a QCD vacuum dominated by instantons \cite{latins}.
\begin{figure}[!ht]
  %\hfill
  %\begin{minipage}[h]{.45\textwidth}
   % \begin{center}  
      \includegraphics[width=\columnwidth]{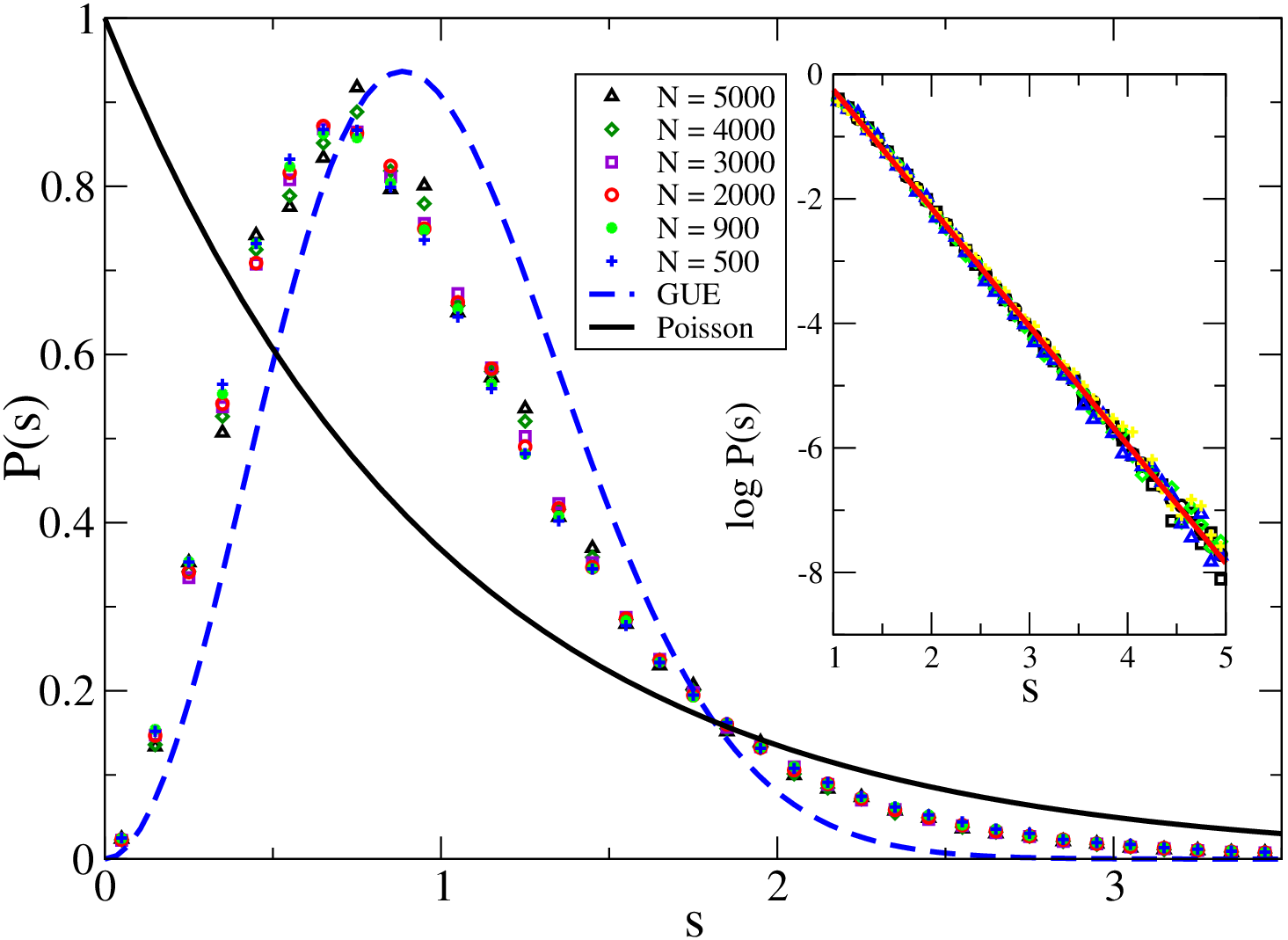}
      \caption{Level spacing distribution $P(s)$ in the bulk for the quenched
               ILM at $T=200$ MeV for different volumes.
               The inset shows $\log P(s)$ for the tail of the data.
               The best fit (solid line) corresponds to a slope $-1.64$.}
      \label{nf0ps1}
    %\end{center}
  %\end{minipage}
  %\hfill
     \end{figure}
%  \begin{minipage}[h]{.45\textwidth}
%    \begin{center}  
        \begin{figure}[b]
      \includegraphics[width=\columnwidth]{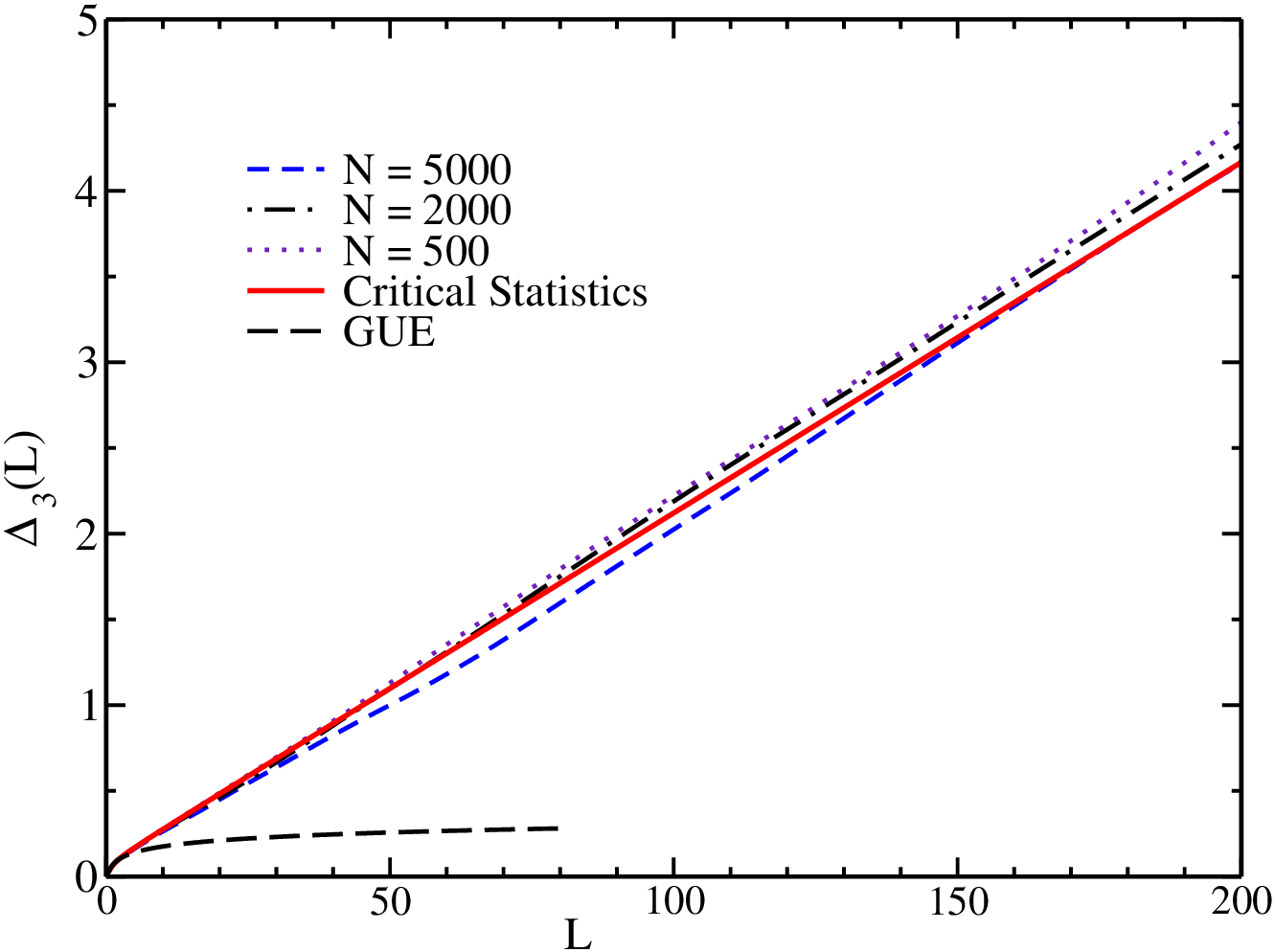}
      \caption{Spectral rigidity 
in the bulk for the quenched ILM at $T=200$ MeV
 for different volumes.
 The result has very little size dependence and agrees well with
 the prediction of critical statistics (solid line).}
      \label{nf0d3}
%    \end{center}
%  \end{minipage}
%  \hfill
\end{figure}

\subsection{Technical details of the numerical simulation}

The ILM  partition function for $N_f$ quark flavors with masses $m_k$
is given by
\be
Z_{\rm inst} = \int D\Omega  \; e^{-S_{\rm YM}}\;
 \prod_{k=1}^{N_f} {\det}(\slashiii{D} + m_k),
\label{zinst}
\ee
where the integral is over the positions, sizes and orientations of the
instantons
and $S_{\rm YM}$ is the Yang-Mills action.
The fermion determinant is evaluated in the space
of the fermionic zero modes of the instantons. 
For further discussion of this partition function we refer to \cite{SS97}.
We just mention that we use the standard phenomenological value of the
instanton density $N/V = 1\,\mathrm{fm}^{-4}$. For the sake 
of simplicity we have kept this density fixed as the temperature is
increased. We justify this approximation based on the 
 fact that even for temperatures close to the chiral phase
transition the instanton density is still sizable 
(around $0.6\,\mathrm{fm}^{-4}$ \cite{SS97}).
Indeed in previous simulations the drop in the density was ruled out as
 the physical mechanism leading to the chiral phase transition \cite{SS97}.

All units in the ILM are typically given in terms of the QCD scale parameter
$\Lambda$ which we have simply set to 200 MeV.
This choice is sufficient for our purposes since we are not concerned with
making quantitative predictions about the position of the QCD chiral
phase transition, but instead are interested in more general features of
the transition that we don't expect to be very sensitive to a fine tuning
of the parameters of the ILM.
We therefore use the
ILM as a qualitative model for QCD at
temperatures around the chiral phase transition.
We also stress that it provides a reasonable
 description of \SB{} and many hadronic correlation functions both at zero
 and at finite temperature \cite{SSV}.
Furthermore the ILM has modest computational requirements and allows
 us to go to fairly large volumes and get good statistics. 

In this paper we present results for both the quenched $N_f = 0$ and
the unquenched $N_f=2$ cases ($N_f=2$ with $m_u = m_d = 0)$.
The partition function, equation (\ref{zinst}), is evaluated using a standard
Metropolis algorithm.
In the quenched case we performed between 2000 to 10000 measurement
sweeps for each set of parameters after allowing around 1000 sweeps
for thermalization.
Results are presented for ensembles of up to $N=5000$ instanton and
anti-instantons.
In the unquenched case we were only able to
investigate ensembles of up to $N=700$ instantons and anti-instantons.
We also discarded the first 1000 sweeps in each simulation and did between
1000 to 5000 sweeps per ensemble.
Thermodynamic properties of the model have already been addressed in
\cite{SS97} and will not be discussed here.
We will focus mainly on observables 
such as level statistics and eigenvector scaling properties of the
 QCD Dirac operator which are especially well suited
 for studying Anderson localization effects \cite{mirlin}.
All the spectral correlators are calculated from the unfolded spectrum.
This procedure scales the eigenvalues so that the spectral 
density on a spectral window comprising
several level spacings is unity but 
it does not remove the small fluctuations about the mean density that
provide fundamental information about the system.

Finally we recall that in the ILM the effective disorder parameter is the
temperature since we have fixed the density of instantons. One 
of the most challenging tasks of the numerical calculation is to 
accurately locate the temperature 
at which the AT occurs. In all cases in the 
paper this temperature has been estimated by 
the finite size scaling method introduced in \cite{sko} 
in the context of disordered systems.
In essence this method consists of computing a spectral correlator, such as the
level spacing distribution $P(s)$ or the number variance $\Sigma^2(\ell)$, 
for different sizes and then finding the temperature at which it becomes size
independent.

\section{Localization in the Instanton Liquid Model: The quenched case}

In this section we present results for the quenched ILM
at nonzero temperature.
Without dynamical fermions there is not technically a chiral phase
transition since there is no chiral symmetry.
However one can in principle study
the ``quenched'' quark condensate in a purely gluonic background 
 through the Banks-Casher relation \cite{bank}.
However, in the quenched ILM, the spectral density seems to diverge close
 to the origin \cite{chisbinst}
 (similarly to recent lattice results with overlap fermions \cite{kiskis})
 thus suggesting a likewise divergent ``quenched'' quark condensate.
In view of these facts, the transition to localization (see below) 
 we have observed in the quenched QCD Dirac operator at finite temperature
 both close to origin and in the bulk of the spectrum 
 can't be linked directly to either of these phenomena. 
We still consider the quenched ILM results to be of interest 
 as we have found a very clear example of a mobility edge
 with features strikingly similar to those of a 3D disordered system
 at the AT.

First we examine the properties of eigenstates in the bulk of the spectrum.  
This region is not directly related to the chiral transition
however it is a very clear example of a mobility edge
 in the ILM. Moreover since it %effects or affects?
affects a larger range of eigenmodes we can get better
 statistics.
By using the finite scaling method introduced in \cite{sko} we have observed a
 mobility edge separating localized from delocalized eigenstates
 in the range $T \sim$ 150 -- 250 MeV. As the temperature
decreases the location of the mobility edge moves to the end of the
spectrum. For $T < 170$ MeV the results are less reliable since the
mobility edge is located almost in the end of the spectrum where
truncation effects are larger. 
 Unlike the unquenched case, where the
first temperature to study is in principle dictated by the restoration
temperature, we do not have a special reason to choose a specific
temperature to investigate the AT.
 
 Following the literature in
disordered systems we have investigated the temperature $T \sim 200$ MeV
such that the mobility edge is located around the center of the
spectrum, namely, if the spectrum spans a spectral window $[0,e_f]$, the center 
corresponds with a region around $\sim e_f/2$. 

The level statistics in this spectral region 
have all the signatures of an AT including a
spectrum that is scale invariant to high degree. As observed in
 figures \ref{nf0ps1} and \ref{nf0d3} both short range statistics such as
the level spacing distribution, $P(s)$, and long range 
 correlators such as the spectral rigidity
  $\Delta_{3}(\ell)=\frac{2}{\ell^4}\int_{0}^{\ell}(\ell^3-2\ell^2x+x^3)\Sigma^{2}(x)dx$
do not depend on the system size
 for volumes (number of instantons) ranging from $N=500$ to $N=5000$.
 Level repulsion is still present as for a 
 metal but the spectral rigidity (see figure \ref{nf0d3}) is asymptotically
 linear with a slope that corresponds to $\chi \sim
 0.29 \pm 0.02$ in fair agreement with the value for a 3D disordered system at
 an AT of $\chi \sim 0.27\pm 0.02$ \cite{chiat}. Furthermore the 
 exact form of the spectral rigidity follows closely the prediction 
 of critical statistics, equation (31) in \cite{ant4} with $h$,
 a free parameter, set to a value of $0.62$.
 %As observed in figure \ref{nf0ps}
%the level statistics outside the critical region are
% RMT like in the region between the origin and the mobility edge
% and Poisson like between the mobility edge and the end of the spectrum. 
 
We now investigate the multifractality of the
eigenstates by looking at the scaling of
 $P_q=\int d^dr |\psi({\bf r})|^{2q}\propto L^{-D_q(q-1)}$,
 where $D_q$ is a set of different exponents describing the
 transition and $L \propto N^{1/3}$ is the spatial system size.
The second moment $P_2$ is usually referred to as the inverse
participation ratio (IPR).

In order not to consider eigenstates
outside the critical region we have taken only $10\%$ of the
eigenvectors around the center of the spectrum for $T=200$ MeV. 
By fitting $P_2$ (also known as the inverse participation ratio or IPR)
for different volumes, 
 we have obtained a value of $D_2 \sim 1.5 \pm 0.1$ in 
 agreement with that of a 3D disordered system at the AT \cite{mirlin}.

Our spectral and eigenfunction analysis shows that at $T=200$ MeV there is mobility edge
 around the central part of the spectrum in the quenched ILM.
%In conlusion, level statistics have all the signatures of an Anderson transition:
% scale invariance, level repulsion and sub-Poisson spectral rigidity.
We have also observed similar critical features in the
 region close to the origin at temperatures around 110--140 MeV.
%A
Here we expect the appearance of the mobility edge to be related to \SB,
However the analysis is complicated by the accumulation of very small
eigenvalues typical of the quenched approximation in QCD.
Additionally numerical results are less conclusive due 
to poor statistics. Despite these limitations we have also observed
 critical statistics and multifractal eigenstates with $D_2 \sim 0.8
\pm 0.1$ (more sparse than in the bulk).
%More details of this region will be given elsewhere \cite{aj2}.
%Instead we present results near the origin for two massless flavors 
%where the accumulation of very small eigenmodes is suppressed and
%the analysis is cleaner.

\begin{figure}[!ht]
%  \hfill
%  \begin{minipage}[t]{.45\textwidth}
%    \begin{center}  
\includegraphics[width=\columnwidth,clip,angle=0]{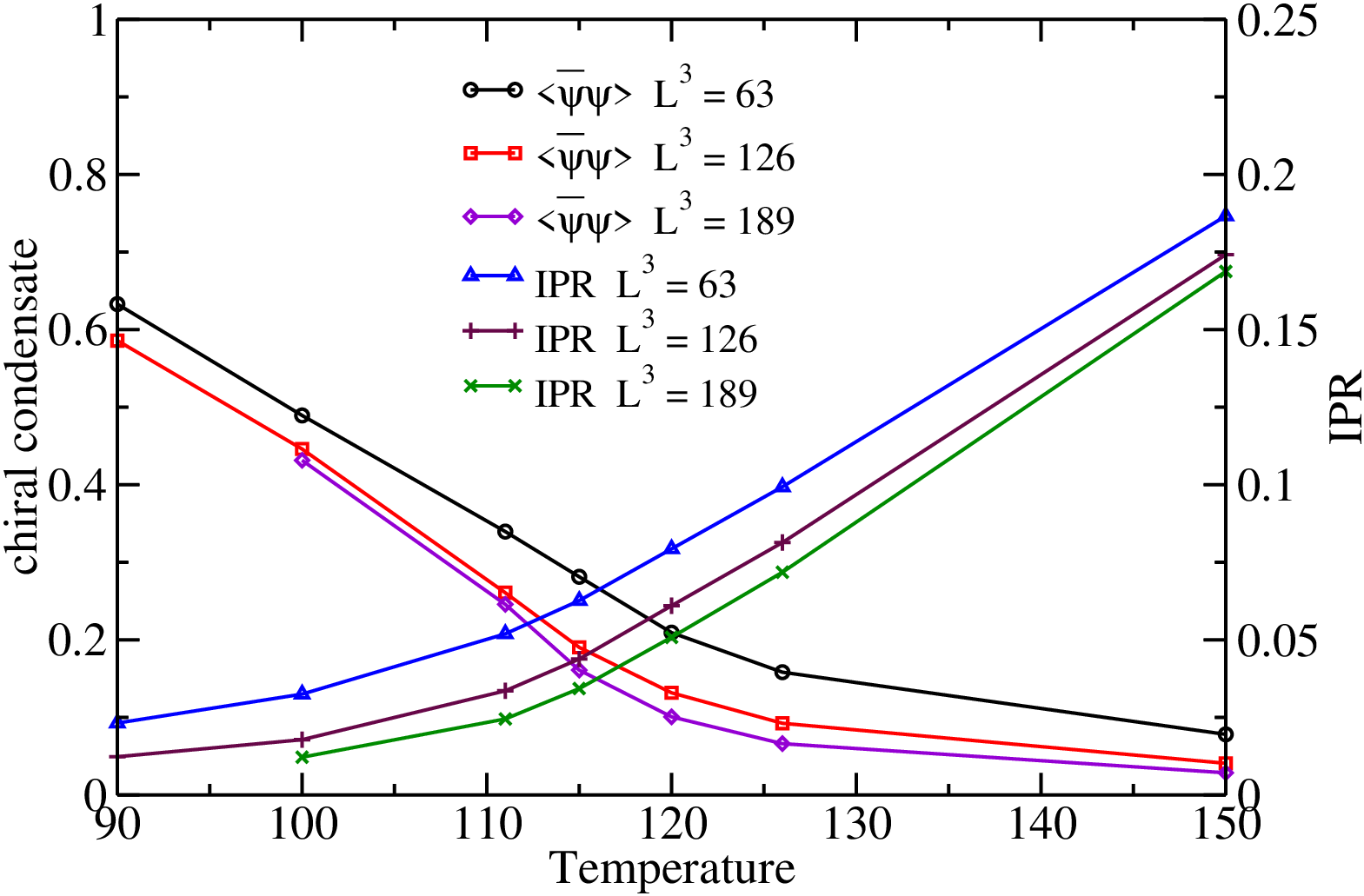}
\caption{Chiral condensate and inverse participation ratio (IPR) of the lowest eigenmode for the ILM with two massless quarks.}
\label{cciprnf2m0}
%    \end{center}
%  \end{minipage}
%  \hfill
%  \begin{minipage}[t]{.45\textwidth}
%    \begin{center}  
\end{figure}
\begin{figure}[!ht]
\includegraphics[width=\columnwidth,clip,angle=0]{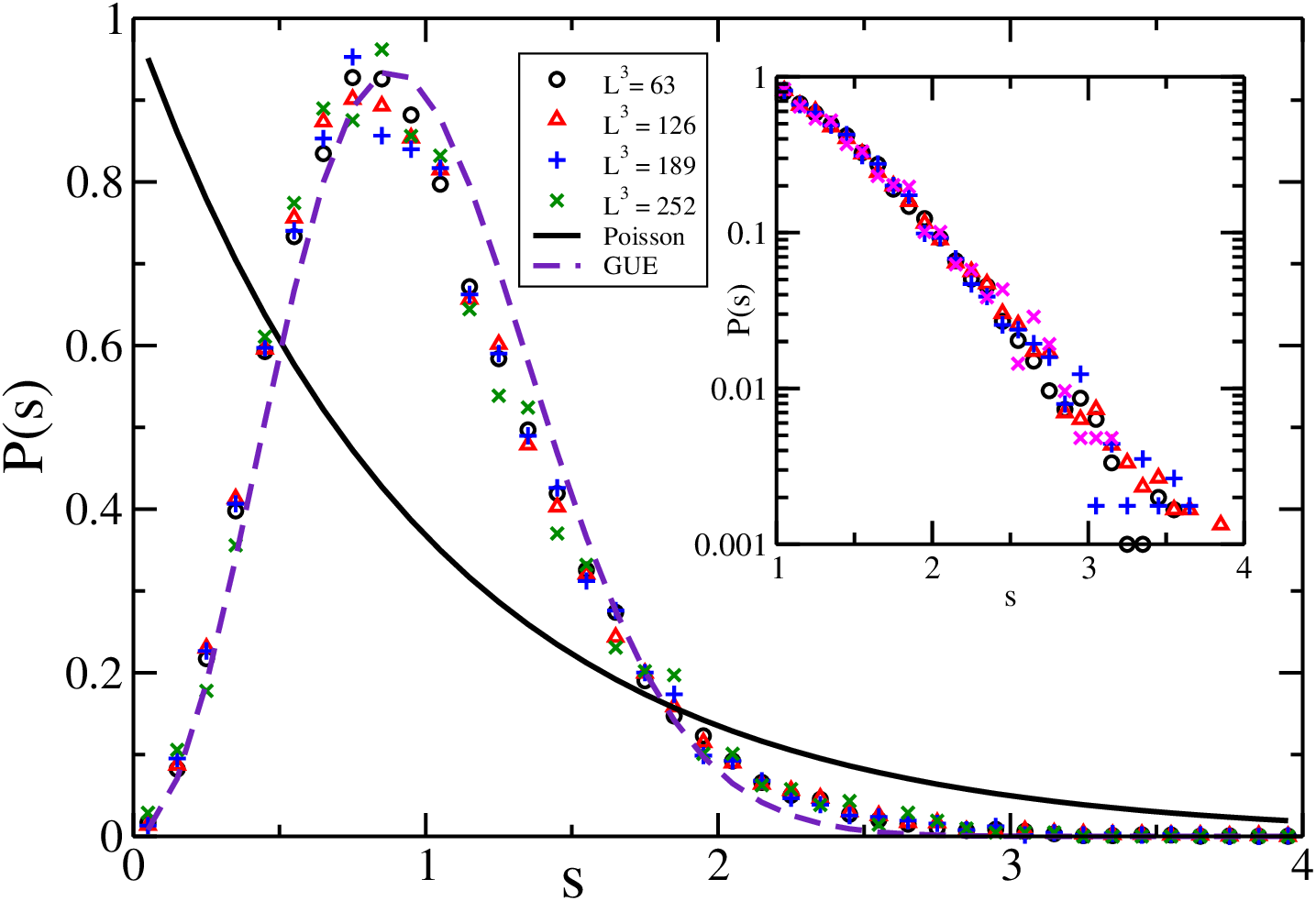}
\caption{Level spacing distribution $P(s)$ at the origin for the unquenched ILM
 at $T=115$ MeV for different system sizes.}
\label{psnf2m0}
%    \end{center}
%  \end{minipage}
%  \hfill
\end{figure}

\section{Chiral restoration and Anderson transition: The unquenched case}

We now study eigenvalues and eigenvectors of the QCD
Dirac operator in an ensemble with two massless quark flavors.
In this case the chiral condensate associated with \SB{} is finite.
Around the critical temperature, the condensate approaches zero signaling
 the chiral phase transition.
Since we have two massless flavors we expect to find a
 second order chiral phase transition.

We focus our attention on the spectral 
 region close to origin since our aim is to relate the chiral phase transition 
 with a transition to localization (AT) of the QCD Dirac operator eigenmodes. 
By looking at the chiral condensate versus temperature for a range
of system sizes (figure \ref{cciprnf2m0}) we see that the condensate
%A1 zero missing 
approaches zero around temperatures of 115 -- 120 MeV in a manner consistent
with a second order phase transition.
At the same time the condensate is falling we see the IPR of the lowest
 eigenmode begin to rise signaling a transition to localization. 
This finding strongly suggests that both phenomena are
 intimately related as was first conjectured in Ref.\cite{diakonov}.

Next we examine the nature of the localization transition.
First we locate, by using the finite size scaling method \cite{sko},
%A1 -120 
 the AT for the lowest lying eigenmodes at $T \sim$ 115 -- 120 MeV.
At this temperature we see that eigenvalue statistics such as the
level spacing distribution (figure \ref{psnf2m0}) are nearly scale
invariant, there is level repulsion but the tail of $P(s)$ is exponential as at the AT.
We also looked at the scaling of eigenmode moments 
$P_q$ 
in the lowest eigenmode
 at $T=120$ MeV for system sizes ranging from $L^3=$63--252.
This yielded a set of fractal dimensions
%A1
$D_2=1.3\pm0.2$, $D_3=0.9\pm0.2$, $D_4=0.7\pm 0.3$ and $D_5=0.7\pm0.3$
%$D_2=2.2\pm0.2$, $D_3=1.9\pm0.3$, $D_4=1.6\pm0.4$ and $D_5=1.3\pm0.5$
%A1 eigenstates instead system
%showing that the eigenstates are multifractal.
Although higher volumes would be desirable to 
fully confirm the multifractal scaling of the eigenstates, it is
encouraging that the IPR of comparatively similar temperature
 ($T \sim 100$ MeV) is comparable to the one for a fully metallic sample.
%{\bf I don't think this is really that clear, maybe T=100 is a better example} A1 OK
We also mention that we observed a
 mobility edge on the bulk of the spectrum at a slightly higher
 temperature with properties similar to the quenched case.

At this point a natural question to address is  
 whether the results here reported concerning the Anderson transition 
 in the ILM may also be present in more 
 realistic models of the strong interactions.  The answer, at least for lattice QCD, seems to be positive \cite{aj3} though  larger volumes and better statistics 
are still needed to further clarify the role of localization in QCD.
 In this final section we analyze why an Anderson transition is expected in the ILM and to what extent these arguments may also be valid  in lattice QCD.

As is known, unlike the zero temperature case where the decay is power-law, 
%A1 caloron defined
the fermionic zero modes in the field of an instanton at nonzero temperature
 (usually called a caloron)
%A1 are by it is, not sure
have an exponential tail $e^{-rT}$ in the spatial directions and are
oscillatory in the time direction \cite{SS97}.
This suggests that the overlap among different zero modes is essentially
 restricted to nearest neighbors in the spatial directions.
However, in the time direction different zero modes strongly overlap
 due to the oscillatory character of the eigenmodes.
This situation strongly resembles a 4D disordered conductor 
 in the tight binding approximation (only nearest neighbor hopping) 
 with one dimension (time) much smaller than the rest so the system can be
 considered effectively three dimensional.
It is well established that such a system may undergo an AT depending on the disorder
 strength.
In our case the role of disorder is played by temperature since the
wavefunction decays as $\sim e^{-rT}$. 
From this discussion it is natural to find an 
Anderson transition in the ILM for a particular value of the 
temperature as first suggested in Ref. \cite{diakonov}.

The principal ingredient to reach the AT is thus an exponential decay
 of the eigenmodes explicitly depending on the temperature together with the 
 possibility to tune the effective range of the exponential in order to 
 reach the transition region.
Any theory with these features very likely 
 will undergo an AT for some value of the parameters (see Ref. \cite{aj2} for a more detailed discussion).
It is therefore quite possible that the QCD vacuum has similar properties.
Even if the objects responsible for localization are not the classical
%A1 \ref should be \cite? caloron defined previously
 instantons 
 this scenario could still be realized
 if the interactions behave in a similar way.
%A1 memoved n in scenario 
We refer to \cite{aj2} for a more detailed 
 account of this interesting issue.    

In summary we have found that in the unquenched ILM the chiral
restoration transition occurs around the same temperature as the localization
transition close to the origin. This indicates that both phenomena are closely
related.
From a physical point of view this is
not surprising since the appearance of the condensate is deeply linked \cite{bank}
with the delocalization of the zero modes in the instanton vacuum due to
long range instanton induced interactions.  
It will be very interesting to repeat these same studies in lattice
simulations where the chiral transition also coincides with a
deconfinement transition and additionally to see if the detailed structure
of the topological objects
affects the nature of the localization transition.


\begin{thebibliography}{9}
%\vspace{-15mm}

\bibitem{shuryak} E. Shuryak, \emph{Nucl. Phys.} {\bf B203} (1982) 93,116,140.

\bibitem{chisbinst}
%\bibitem{diakonov1}
 D. Diakonov and P. Petrov, \emph{Nucl. Phys.} {\bf B272}
 (1986) 457; \emph{Phys. Lett.} {\bf B147} (1984) 351; hep-ph/9602375;
%\bibitem{VO}
 J.C. Osborn and J.J.M. Verbaarschot, \emph{Phys. Rev. Lett.}
 {\bf 81} (1998) 268; \emph{Nucl. Phys.} {\bf B525} (1998) 738;
%\bibitem{zahed}
 R.A. Janik, M.A. Nowak, G. Papp, and I. Zahed,
 \emph{Phys. Rev. Lett.} {\bf 81} (1998) 264.

\bibitem{polyakov} A. Belavin, A. Polyakov, A. Schwartz and Y. Tyupkin,
 \emph{Phys. Lett.} {\bf 59} (1975) 85;
 G. 't Hooft, \emph{Phys. Rev. Lett.} {\bf 37} (1976) 8.

\bibitem{anderson} P.W. Anderson, \emph{Phys. Rev.} {\bf 109} (1958) 1492.
\bibitem{diakonov1} D. Diakonov and V. Petrov,
 \emph{Nucl. Phys.} {\bf B245} (1984) 259.
\bibitem{aoki} H. Aoki, \emph{J. Phys.} {\bf 16C}, (1983) L205;
 F. Wegner, \emph{Z. Phys. B} {\bf 36} (1980) 209.
\bibitem{diakonov}D. Diakonov, Lectures at the Enrico Fermi School in Physics, Varenna, hep-ph/9602375.
\bibitem{kravtsov97} V.E. Kravtsov and K.A. Muttalib, \emph{Phys. Rev. Lett.}
 {\bf 79} (1997) 1913.
\bibitem{kiskis} J. Kiskis and R. Narayanan,
 \emph{Phys.Rev.} D {\bf 64} (2001) 117502.

\bibitem{SSV} T. Shafer, E. Shuryak and J.J.M. Verbaarschot,
 \emph{Nucl. Phys.} {\bf B412} (1994) 143.
\bibitem{sko}B.I. Shklovskii, et.al.,
% B. Shapiro, B.R.~Sears,P.~Lambrianides and H.B.~Shore, 
\emph{Phys. Rev. B} {\bf 47} (1993) 11487.
\bibitem{aj2}A. M. Garcia-Garcia, J. C. Osborn, \emph{Nucl.Phys.}{ \bf A770} (2006) 141.
\bibitem{chiat} D. Braun, G. Montambaux and M. Pascaud,
 \emph{Phys. Rev. Lett.} {\bf 81} (1998) 1062.
\bibitem{latins} M.-C. Chu, et. al.,
 \emph{Phys. Rev.} D {\bf 49} (1994) 6039;
 C. Michael and P.S. Spencer, \emph{Phys. Rev.} D {\bf 52} (1995) 4691.
\bibitem{SS97} T. Sch\"afer and E. Shuryak, \emph{Rev. Mod. Phys.}
 {\bf 70} (1998) 323.
\bibitem{bank} T. Banks and A. Casher,
 \emph{Nucl. Phys.} {\bf B169} (1980) 103.
\bibitem{ant4} A.M. Garcia-Garcia and J.J.M. Verbaarschot,
 \emph{Phys. Rev. E} {\bf 67} (2003) 046104.

\bibitem{mirlin} T. Ohtsuki and T. Kawarabayashi,
 \emph{J. Phys. Soc. Jpn.} {\bf 66} (1997) 314. 
\bibitem{aj3}A. M. Garcia-Garcia, J. C. Osborn, \emph{Phys. Rev. D}, {\bf 75} (2007) 034503.
\end{thebibliography}
\end{document}